\documentclass[sigconf]{aamas}  
\usepackage{balance}  

\usepackage{booktabs}
\usepackage{subfigure}
\usepackage{graphicx}
\usepackage{algorithm}
\usepackage{algpseudocode}
\usepackage[T1]{fontenc}
\usepackage[latin9]{inputenc}
\usepackage{float}
\usepackage{amsmath}
\usepackage{amssymb}
\usepackage{stackrel}
\usepackage{inputenc} 
\usepackage{url}            
\usepackage{amsfonts}       
\usepackage{nicefrac}       
\usepackage{microtype}      
\usepackage{multirow}
\usepackage{tabularx}

\setcopyright{ifaamas}  
\acmDOI{}  
\acmISBN{}  
\acmConference[AAMAS'19]{Proc.\@ of the 18th International Conference on Autonomous Agents and Multiagent Systems (AAMAS 2019)}{May 13--17, 2019}{Montreal, Canada}{N.~Agmon, M.~E.~Taylor, E.~Elkind, M.~Veloso (eds.)}  
\acmYear{2019}  
\copyrightyear{2019}  
\acmPrice{}  

\providecommand{\tabularnewline}{\\}
\floatstyle{ruled}
\allowdisplaybreaks
\begin{document}

\title[Optimal Control of Complex Systems through Variational Inference  with DEDP]{Optimal Control of Complex Systems through\texorpdfstring{\\}{} Variational Inference with a Discrete Event Decision Process}

\author{Fan Yang}
\affiliation{%
  \institution{University at Buffalo}
  \city{Buffalo} 
  \state{New York} 
  \postcode{14226}
}
\email{fyang24@buffalo.edu}
\author{Bo Liu}
\affiliation{%
  \institution{Auburn University}
  \city{Auburn} 
  \state{Alabama} 
  \postcode{36849}
}
\email{boliu@auburn.edu}
\author{Wen Dong}
\affiliation{%
  \institution{University at Buffalo}
  \city{Buffalo} 
  \state{New York} 
  \postcode{14226}
}
\email{wendong@buffalo.edu}

\begin{abstract}
Complex social systems are composed of interconnected individuals
whose interactions result in group behaviors. Optimal control of a
real-world complex system has many applications, including road traffic management, epidemic prevention, and information dissemination.
However, such real-world complex system control is difficult to achieve
because of high-dimensional and non-linear system dynamics, and the
exploding state and action spaces for the decision maker. Prior methods
can be divided into two categories: simulation-based and analytical approaches.
Existing simulation approaches have high-variance in Monte Carlo integration, and the
analytical approaches suffer from modeling inaccuracy. We adopted
simulation modeling in specifying the complex dynamics of a complex
system, and developed analytical solutions for searching optimal strategies
in a complex network with high-dimensional state-action space. To
capture the complex system dynamics, we formulate the complex social network
decision making problem as a discrete event decision process. To address
the curse of dimensionality and search in high-dimensional state
action spaces in complex systems, we reduce control of a complex system
to variational inference and parameter learning, introduce Bethe entropy
approximation, and develop an expectation propagation algorithm. Our
proposed algorithm leads to higher system expected rewards, faster
convergence, and lower variance of value function in a real-world
transportation scenario than state-of-the-art analytical and sampling
approaches.
\end{abstract}

\maketitle



\section{Introduction}
A complex social system is a collective system composed of a large
number of interconnected entities that as whole exhibit properties
and behaviors resulted from the interaction of its individual parts
\cite{page2015sociologists}. Achieving optimal control of a real-world
complex social system is important and valuable. For example, in an
urban transportation system, a central authority wants to reduce the
overall delay and the total travel time by controlling traffic signals
using traffic data collected from the Internet of Things \cite{balaji2010urban}.
In an epidemic system, an optimal strategy is pursued to minimize
the expected discounted losses resulting from the epidemic process
over an infinite horizon, with specific aims such as varying the birth
and death rates \cite{lefevre1981optimal} or early detection of epidemic
outbreaks \cite{izadi2007optimizing}. In the sphere of public opinion,
components such as information consumers, news media, social websites,
and governments aim to maximize their own utility functions with optimal
strategies. A specific objective in the public opinion environment
is tackling fake news to create a clearer, more trusted information
environment \cite{allcott2017social}. These scenarios exemplify the
enormous potential applications of a versatile optimal control framework
for general complex systems.

However, key characteristics of complex systems make establishing such
a framework very challenging. The system typically has a large number
of interacting units and thus a high-dimensional state space. State
transition dynamics in complex systems are already non-linear, time-variant,
and high-dimensional, and then these frameworks must account for additional
control variables. Prior research in decision making for complex social
systems has generally gone in one of two directions: a simulation
or analytical approach \cite{li2014survey}. Simulation approaches
specify system dynamics through a simulation model and develop sampling-based
algorithms to reproduce the dynamic flow \cite{wang2010parallel,preciado2013optimal,yang2018integrating,yang2018predicting}. These approaches can capture the microscopic dynamics of a complex
system with high fidelity, but have a high variance and are time-consuming \cite{li2016traffic}.
Analytical approaches instead formulate the decision-making problem
as a constrained optimization problem with an analytical model through specifying the macroscopic state transitions
directly, deriving
analytical solutions for optimizing strategies \cite{farajtabar2015coevolve,timotheou2015distributed}.
These approaches can provide a robust solution with
less variance, but are applicable only to scenarios with small
state spaces \cite{kautz2004learning}, or cases with low resolution
intervention \cite{he2010optimal}, due to modeling costs and errors
\cite{li2016traffic}. The above research points to a new direction in
research opportunities that combines the ability to capture system
dynamics more precisely from simulation approaches with the benefit
of having less variance and being more robust from analytical approaches.
In this paper, we adopted simulation modeling in specifying the
dynamics of a complex system, and developed analytical solutions for
searching optimal strategies in a complex network with high-dimensional
state-action space specified through simulation modeling.

We formulate the problem of decision making in a complex system as
a discrete event decision process (DEDP), which identifies the decision
making process as a Markov decision process (MDP) and introduces a
discrete event model \textemdash{} a kind of simulation model \cite{law1991simulation}
\textemdash{} to specify the system dynamics. A discrete event model
defines a Markov jump process with probability measure on a sequence
of elementary events that specify how the system components interact
and change the system states. These elementary events individually effect
only minimal changes to the system, but in sequence together are powerful
enough to induce non-linear, time-variant, high-dimensional behavior.
 Comparing with an MDP which specifies the system transition dynamics
of a complex system analytically through a Markov model, a
DEDP describes the dynamics more accurately through a discrete event
model that captures the dynamics using a simulation process over the
microscopic component-interaction events. We will demonstrate this
merit through benchmarking with an analytical approach based on a
MDP in the domain of transportation optimal control.

To solve a DEDP analytically, we derived a duality theorem that recasts
optimal control to variational inference and parameter learning, which
is an extension of the current equivalence results between optimal
control and probabilistic inference \cite{liu2012belief,toussaint2006probabilistic}
in Markov decision process research. With this duality, we can include
a number of existing probabilistic-inference and parameter-learning
techniques, and integrate signal processing and decision making into
a holistic framework. When exact inference becomes intractable, which
is often the case in complex systems due to the formidable state space,
our duality theorem implies the possibility of introducing recent
approximate inference techniques to infer complex dynamics. The method
in this paper is an expectation propagation algorithm, part of a family
of approximate inference algorithms with local marginal projection.
We will demonstrate that our approach is more robust and has less
variance in comparison with other simulation approaches in the domain
of transportation optimal control.

This research makes several important contributions. First, we formulate
a DEDP \textemdash{} a general framework for modeling complex system
decision-making problems \textemdash{} by combining MDP and simulation
modeling. Second, we reduce the problem of optimal control to variational
inference and parameter learning, and develop an approximate solver
to find optimal control in complex systems through Bethe entropy approximation
and an expectation propagation algorithm. Finally, we demonstrate
that our proposed algorithm can achieve higher system expected rewards,
faster convergence, and lower variance of value function within a
real-world transportation scenario than even state-of-the-art analytical
and sampling approaches.

\section{Background}

In this section, we review the complex social system, the discrete event model, the Markov decision
process, and the variational inference framework for a probabilistic
graphical model. 

\subsection{Complex Social System}

A complex social system is a collective system composed of a large
number of interconnected entities that as whole exhibit properties
and behaviors resulted from the interaction of its individual parts.
A complex system tends to have four attributes: Diversity, interactivity,
interdependency, and adaptivity \cite{page2015sociologists}. Diversity
means the system contains a large number of entities with various
attributes and characteristics. Interactivity means the diverse entities
interact with each other in an interaction structure, such as a fixed
network or an ephemeral contact. Interdependency means the state change
of one entity is dependent on others through the interactions. Adaptivity
means the entities can adapt to different environments automatically. 

In this paper, we temporarily exclude the attribute of adaptivity,
the study of which will be future work. We focus on studying the optimal
control of a complex social system with the attributes of diversity,
interactivity, and interdependency, which leads to a system containing
a large number of diverse components, the interactions of which lead
to the states change. Examples of complex social systems include the
transportation system where the traffic congestions are formed and
dissipated through the interaction and movement of individual vehicles,
the epidemic system where the disease is spread through the interaction
of different people, and the public opinion system where people's
minds are influenced and shaped by the dissemination of news through
social media.

\subsection{Discrete Event Model}


A discrete event model defines a discrete event process, also called
a Markov jump process. It is used to specify complex system dynamics
with a sequence of stochastic events that each changes the state only
minimally, but when combined in a sequence induce complex system evolutions.
Specifically, a discrete event model describes the temporal evolution
of a system with $M$ species $\mathcal{X}=\{X^{(1)},X^{(2)},\cdots,X^{(M)}\}$
driven by $V$ mutually independent events parameterized by rate coefficients
$\mathbf{c}=(c_{1},\dots,c_{V})$. At any specific time $t$, the
populations of the species are $x_{t}=(x_{t}^{(1)},\dots,x_{t}^{(M)})$. 

A discrete event process initially in state $x_{0}$ at time $t=0$
can be simulated by: (1) Sampling the event $v\in\left\{ \emptyset,1,\cdots,V\right\} $
according to categorical distribution $v\sim(1-h_{0},h_{1},\dots,h_{V})$,
where $h_{v}(x,c_{v})=c_{v}\prod_{m=1}^{M}g_{v}^{(m)}(x_{t}^{(m)})$
is the rate of event $v$, which equals to the rate coefficients $c_{v}$
times a total of $\prod_{m=1}^{M}g_{v}^{(m)}(x^{(m)})$ different
ways for the individuals to react, and $h_{0}(x,c)=\sum_{v=1}^{V}h_{v}(x,c_{v})$
the rate of all events. The formulation of $h_{v}(x,c_{v})$ comes
from the formulations of the stochastic kinetic model and stochastic
petri net \cite{xu2016using,wilkinson2011stochastic}. (2) Updating
the network state deterministically $x\leftarrow x+\Delta_{v}$, where
$\Delta_{v}$ represents how an event $v$ changes the system states,
until the termination condition is satisfied. In a social system,
each event involves only a few state and action variables. This generative
process thus assigns a probabilistic measure to a sample path induced
by a sequence of events $v_{0},\dots,v_{T}$ happening between times
$0,1,\cdots,T$, where $\delta$ is an indicator function. \vspace{-5mm}

\begin{align*}
 & P(x_{0:T},v_{0:T})=p(x_{0})\prod_{t=0}^{T}p(v_{t}|x_{t})\delta_{x_{t+1}=x_{t}+\Delta_{v_{t}}},\\
 & \text{where }p(v_{t}|x_{t})=\begin{cases}
1-h_{0}(x_{t},c), & v_{t}=\emptyset\\
h_{k}(x_{t},c_{k}), & v_{t}=k,
\end{cases}
\end{align*}

The discrete event model is widely used by social scientists to specify
social system dynamics \cite{borshchev2013big} where the system state
transitions are induced by interactions of individual components.
Recent research \cite{yang2017integrating,opper2008variational,fang2017expectation} has
also applied the model to infer the hidden state of social systems,
but this approach has not been explored in social network intervention
and decision making.

\subsection{Markov Decision Process}

A Markov decision process \cite{sutton2011reinforcement} is a framework
for modeling decision making in situations where outcomes are partly
random and partly under the control of a decision maker. Formally,
an MDP is defined as a tuple $\text{MDP}\langle S,A,P,R,\gamma\rangle$,
where $S$ represents the state space and $s_{t}\in S$ the state
at time $t$, $A$ the action space and $a_{t}$ the action taken
at time $t$, $P$ the transition kernel of states such as $P(s_{t+1}|s_{t},a_{t})$,
$R$ the reward function such as $R(s_{t},a_{t})$ \cite{farajtabar2015coevolve}
or $R(s_{t})$ \cite{wiering2004intelligent} that evaluates the immediate
reward at each step, and $\gamma\in[0,1)$ the discount factor. Let
us further define a policy $\pi$ as a mapping from a state $s_{t}$
to an action $a_{t}=\mu(s_{t})$ or a distribution of it parameterized
by $\theta$ \textemdash{} that is, $\pi=p(a_{t}|s_{t};\theta)$.
The probability measure of a $\text{length}-T$ MDP trajectory is
$p(\xi_{T})=p(s_{0})\prod_{t=0}^{T-1}p(a_{t}|s_{t};\theta)p(s_{t+1}|s_{t},a_{t})$,
where $\xi_{T}=(s_{0:T},a_{0:T})$. Solving an MDP involves finding
the optimal policy $\pi$ or its associated parameter $\theta$ to
maximize the expected future reward \textemdash{} $\text{arg max}_{\theta}\mathbb{E}_{\xi}(\sum_{t}\gamma^{t}R_{t};\theta)$.

\begin{figure*}
\begin{minipage}[b]{.48\textwidth}
\begin{centering}
\includegraphics[width=0.9\textwidth]{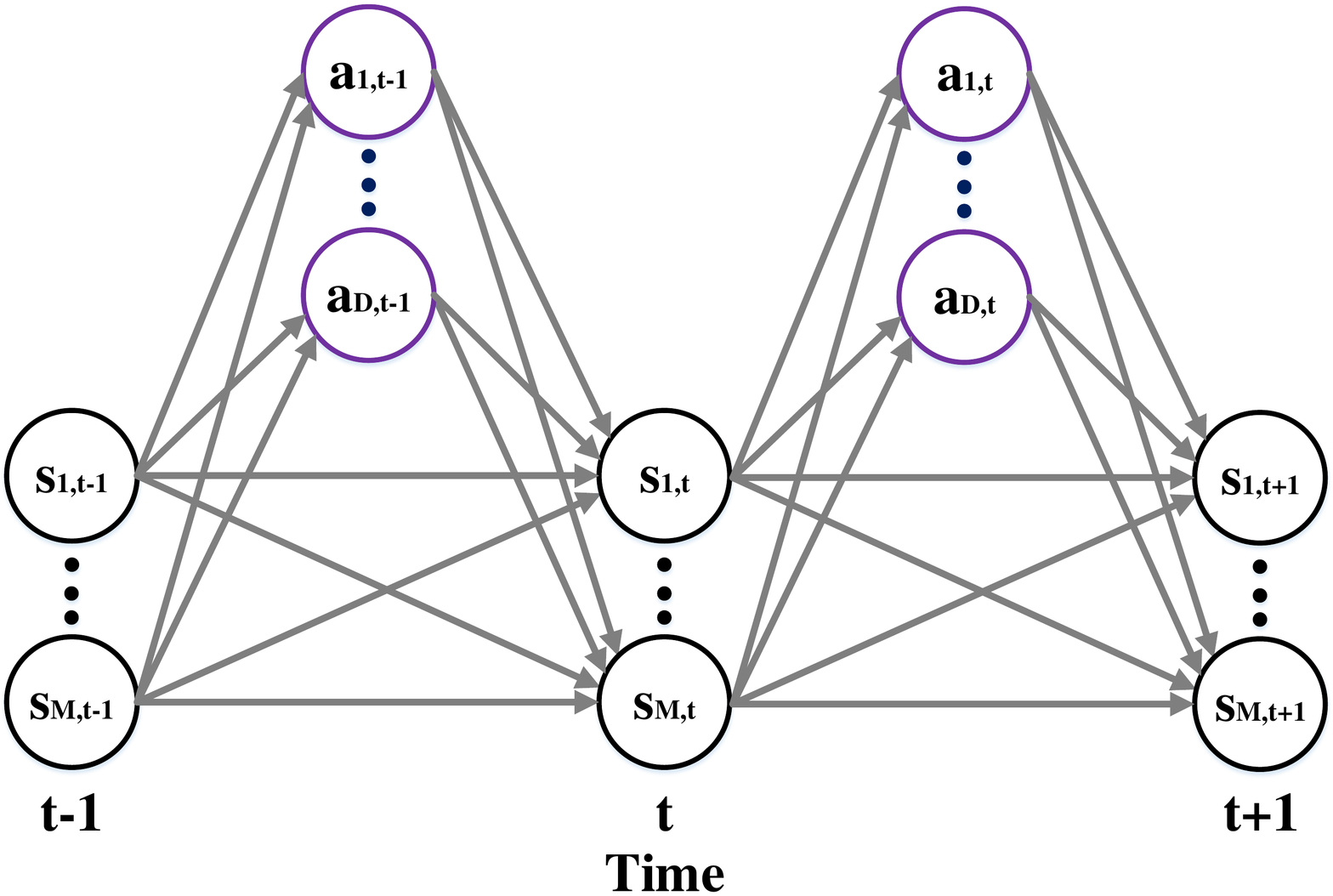}
\par\end{centering}
\vspace{-4mm}
\caption{Graph representation of an MDP\label{fig:MDP}}
\end{minipage}\qquad
\vspace{-1mm}
\begin{minipage}[b]{.48\textwidth}
\begin{centering}
\includegraphics[width=0.9\textwidth]{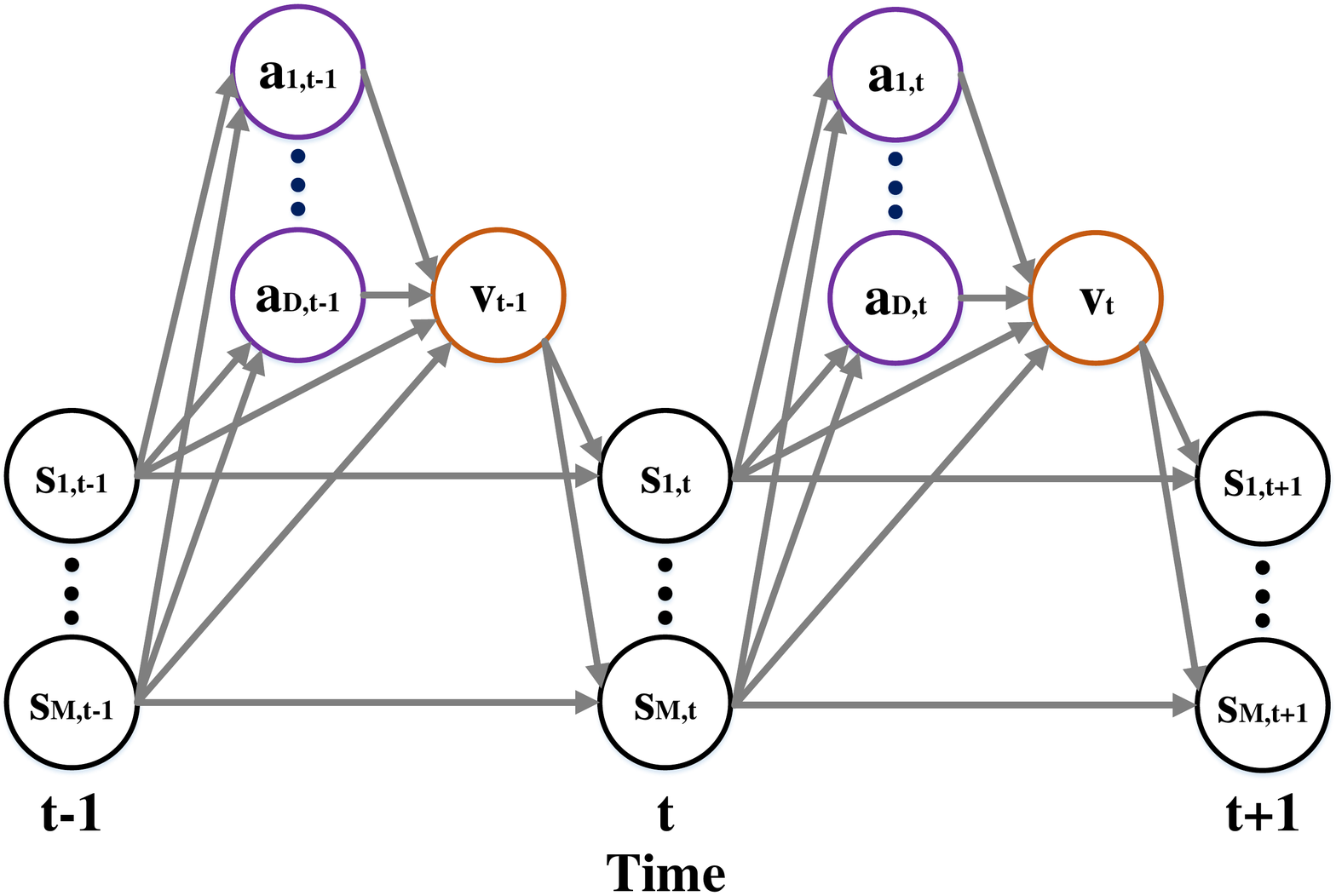}
\par\end{centering}
\vspace{-4mm}
\caption{Graph representation of a DEDP\label{fig:DEDP}}
\end{minipage}
\vspace{-1mm}
\end{figure*}

The graphical representation of an MDP is shown in Figure \ref{fig:MDP},
where we assume that the full system state $s_{t}$ can be represented
as a collection of component state variables $s_{t}=(s_{t}^{(1)},...,s_{t}^{(M)})$,
so that the state space $S$ is a Cartesian product of the domains
of component state $s_{t}^{(m)}$: $S=S^{(1)}\times S^{(2)}\times\cdot\cdot\cdot\times S^{(M)}$.
Similarly, the action variable $a_{t}$ can be represented as a collection
of action variables $a_{t}=(a_{t}^{(1)},...,a_{t}^{(D)})$, and the
action space $A=A^{(1)}\times A^{(2)}\times\cdot\cdot\cdot\times A^{(D)}$.
Here $M$ is not necessarily equal to $D$ because $s_{t}$ represents
the state of each component of the system while $a_{t}$ represents
the decisions taken by the system as a whole. For example, in the
problem of optimizing the traffic signals in a transportation system
where $s_{t}$ represents the locations of each vehicle and $a_{t}$
represents the status of each traffic light, the number of vehicles
$M$ may not necessarily equal to the number of traffic lights $D$.
Usually in complex social systems, the number of individual components
$M$ is much greater than the system decision points $D$. 

Prior research in solving a Markov decision process for a complex
social system could be generally categorized into simulation or analytical approaches. A simulation approach reproduces the dynamic flow through sampling-based method. It describes the state transition dynamics
with a high-fidelity simulation tool such as MATSIM \cite{Horni2016},
which simulates the microscopic interactions of the components and
how these interactions leads to macroscopic state changes. Given current
state $s_{t}$ and action $a_{t}$ at time $t$, a
simulation approach uses a simulation tool to generate the next
state $s_{t+1}$. 

An analytical approach develops analytical solutions to solve a constrained
optimization problem. Instead of describing the dynamics with a simulation
tool, an analytical approach specifies the transition kernel analytically
with probability density functions that describe the macroscopic state
changes directly. Given current state $s_{t}$ and action $a_{t}$,
it computes the probability distribution of the next state $s_{t+1}$
according to the state transition kernel $p(s_{t+1}\mid s_{t},a_{t})$.
However, approximations are required to make the computation tractable.
 For an MDP containing $M$ binary state variables and D binary action
variables, the state space is $2^{M}$, the action space is $2^{D}$,
the policy kernel is a $2^{M}\times2^{D}$ matrix, and the state transition
kernel (fixed action) is a $2^{M}\times2^{M}$ matrix. Since $M$
is usually much larger than $D$ in complex social systems, the complexity
bottleneck is usually the transition kernel with size $2^{M}\times2^{M}$,
the complexity of which grows exponentially with the number of state
variables. Certain factorizations and approximations must be applied
to lower the dimensionality of the transition kernel. 

Usually analytical approaches solve complex social system MDPs approximately
by enforcing certain independence constraints  \cite{sigaud2013markov}.
For example, Cheng \cite{cheng2013variational} assumed that a state variable is only dependent on its neighboring variables. Sabbadin, Peyrard
and Sabbadin \cite{peyrard2006mean,sabbadin2012framework} exploited
a mean field approximation to compute and update the local policies. Weiwei approximated the state transition kernel with differential equations \cite{li2004iterative}. These assumptions introduces
additional approximations that results in modeling errors. In the
next section, we propose a discrete event decision process which reduces
the complexity of the transition probabilities, and which does not
introduce additional independence assumptions.

Two specific approaches of solving an MDP are optimal control \cite{stengel1994optimal}
and reinforcement learning \cite{sutton1998reinforcement}. Optimal
control problems consist of finding the optimal decision sequence
or the time-variant state-action mapping that maximizes the expected
future reward, given the dynamics and reward function. Reinforcement-learning
problems target the optimal stationary policy that maximizes the expected
future reward while not assuming knowledge of the dynamics or the
reward function. In this paper, we address the problem of optimizing
a stationary policy to maximize the expected future reward, assuming
known dynamics and reward function.

\subsection{Variational Inference}

A challenge in evaluating and improving a policy in a complex system
is that the state space grows exponentially with the number of state
variables, which makes probabilistic inference and parameter learning
intractable. For example, in a system with $M$ binary components,
the size of state space $S$ will be $2^{M}$, let alone the exploding
transition kernel. One way to resolve this issue is applying variation
inference to optimize a tractable lower bound of the log expected
future reward through conjugate duality. Variational inference is
a classical framework in the probabilistic graphical model community
 \cite{wainwright2008graphical}. It exploits the conjugate duality
between log-partition function and the entropy function for exponential
family distributions. Specifically, it solves the variational problem
$\text{log}\int\text{exp}\left\langle \theta,\phi(x)\right\rangle dx=\text{sup}_{q(x)}\left\{ \int q(x)\left\langle \theta,\phi(x)\right\rangle dx+H(q)\right\} $,
where $\theta$ and $\phi(x)$ are respectively canonical parameters
and sufficient statistics of an exponential family distribution, $q$
is an auxiliary distribution and $H(q)=-\int dxq(x)\log q(x)$ is
the entropy of $q$. For a tree-structured graphical model (here we
use the simplified notation of a series of $x_{t}$), the Markov property
admits a factorization of $q$ into the product and division of local
marginals $q(x)=\prod_{t}q_{t}(x_{t-1,t})/\prod_{t}q_{t}(x_{t})$.
Substituting the factored forms of $q$ into the variational target,
we get an equivalent constrained optimization problem involving local
marginals and consistency constraints among those marginals, and a
fixed-point algorithm involving forward statistics $\alpha_{t}$ and
backward statistics $\beta_{t}$. There are two primary forms of approximation
of the original variational problem: the Bethe entropy problem and
a structured mean field. The Bethe entropy problem is typically solved
by loopy belief propagation or an expectation propagation algorithm. 

\section{Methodology}

In this section, we develop a DEDP and present a duality theorem that
extends the equivalence of optimal control with probabilistic inference
and parameter learning. We also develop an expectation propagation
algorithm as an approximate solver to be applied in real-world complex
systems. 

\subsection{Discrete Event Decision Process}

The primary challenges in modeling a real-world complex system are
the exploding state space and complex system dynamics. Our solution
is to model the complex system decision-making process as a DEDP. 

The graphical representation of a DEDP is shown in Figure \ref{fig:DEDP}.
Formally, a DEDP is defined as a tuple $\text{DEDP}\langle S,A,\mathcal{V},C,P,R,\gamma\rangle$,
where $S$ is the state space and $s_{t}=\left(s_{t}^{(1)},...,s_{t}^{(M)}\right)\in S$
a M-dimensional vector representing the state of each component at
time $t$, $A$ is the action space and $a_{t}=\left(a_{t}^{(1)},...,a_{t}^{(D)}\right)\in A$
a D-dimensional vector representing the action taken by the system
at time $t$. As before, $M$ is not necessarily equal to $D$, and
$M$ is usually much larger than $D$ in complex social systems. Both
$s_{t}^{(m)}$ and $a_{t}^{(d)}$ could take real or categorical values
depending on the applications.

$\mathcal{V}=\left\{ \emptyset,1,...,V\right\} $ is the set of events
and $v_{t}\in\mathcal{V}$ a scalar following categorical distributions
indicating the event taken at time $t$ and changing the state by
$\Delta_{v_{t}}$. $C$ is the function mapping actions to event rate
coefficients which takes a $D$-dimensional vector as input, and outputs
a $V$-dimensional vector $\boldsymbol{c}=\left(c_{1},...,c_{V}\right)=C(a_{t})$,
and $P$ is the transition kernel of states induced by events $P(s_{t+1},v_{t}|s_{t},a_{t})=p(v_{t}\mid s_{t},a_{t})\delta_{s_{t+1}=s_{t}+\triangle v_{t}}$,
where $p(v_{t}=v\mid s_{t},a_{t})$ represents the probability of
an event $v$ happened at time $t$: \vspace{-4mm}

\[
p(v_{t}=v\mid s_{t},a_{t})=\begin{cases}
h_{v}(s_{t},c_{v}) & \text{if }v\neq\emptyset\\
1-\sum_{v=1}^{V}h_{v}(s_{t},c_{v}) & \text{if }v=\emptyset
\end{cases}
\]

Following the definitions in the discrete event model, $h_{v}(s_{t},c_{v})=c_{v}\cdot\prod_{m=1}^{M}g_{v}^{(m)}(s_{t}^{(m)})$
is the probability for event $v$ to happen, which equals to the rate
coefficients $c_{v}$ times a total of $\prod_{m=1}^{M}g_{v}^{(m)}(s_{t}^{(m)})$
ways that the components can interact to trigger an event. 

The immediate reward function $R$ is a function of system states,
defined as the summation of reward evaluated at each component $R(s_{t})=\sum_{m=1}^{M}R_{t}^{(m)}(s_{t}^{(m)})$
, and $\gamma\in[0,1)$ is the discount factor. We further define
a policy $\pi$ as a mapping from a state $s_{t}$ to an action $a_{t}=\mu(s_{t};\theta)$
or a distribution of it parameterized by $\theta$ \textemdash{} that
is, $\pi=p(a_{t}|s_{t};\theta)$. The parameterized policy can take
any form, such as a lookup table where $\theta$ represents values
of each entries in the table, a Gaussian distribution where $\theta$
represents the mean and variance, or a neural network where $\theta$
represents the network weights. Solving a DEDP involves finding the
optimal policy $\pi$ or its associated parameter $\theta$ to maximize
the expected future reward \textemdash{} $\text{arg max}_{\theta}\mathbb{E}_{\xi}(\sum_{t}\gamma^{t}R_{t};\theta)$.
The probability measure of a $\text{length}-T$ DEDP trajectory with
a stochastic policy is as follows, where $\delta$ is an indicator
function and $\xi_{T}=(s_{0:T},a_{0:T},v_{0:T})$: \vspace{-4mm}

\[
\begin{array}{ll}
p(\xi_{T}) & =p(s_{0})\prod_{t=0}^{T-1}\left(p(a_{t}\mid s_{t})p(v_{t}\mid s_{t},a_{t})\delta_{s_{t+1}=s_{t}+\Delta_{v_{t}}}\right)\end{array}
\]

The probability measure of that with a deterministic policy is this: \vspace{-4mm}

\[
\begin{array}{ll}
p(\xi_{T}) & =p(s_{0})\prod_{t=0}^{T-1}\left(\delta_{a_{t}=\mu(s_{t})}p(v_{t}\mid s_{t},a_{t})\delta_{s_{t+1}=s_{t}+\Delta_{v_{t}}}\right)\\
 & =p(s_{0})\prod_{t=0}^{T-1}\left(p(v_{t}\mid s_{t},\mu(s_{t}))\delta_{s_{t+1}=s_{t}+\Delta_{v_{t}}}\right)\\
 & :=p(s_{0})\prod_{t=0}^{T-1}\left(p(v_{t}\mid s_{t})\delta_{s_{t+1}=s_{t}+\Delta_{v_{t}}}\right)
\end{array}
\]

A DEDP makes a tractable representation of complex system control problem by representing the non-linear and high-dimensional state transition kernel with microscopic events. A vanilla MDP is an intractable representation because the state-action space grows exponentially with the number of state-action variables. In comparison, the description length of a DEDP grows linearly in the number of events. As such, a DEDP greatly reduces the complexity of specifying a complex system control problem through introducing an auxiliary variable (event), and can potentially describe complex and high-fidelity dynamics of a complex system. 

A DEDP could be reduced to an MDP if marginalizing out the events \vspace{-4mm}

\[
\begin{array}{l}
\sum_{v_{0:T}}p(s_{0:T},a_{0:T},v_{0:T})\\
=\sum_{v_{0:T}}p(s_{0})\prod_{t=0}^{T-1}\left(p(a_{t}\mid s_{t})p(v_{t}\mid s_{t},a_{t})\delta_{s_{t+1}=s_{t}+\Delta_{v_{t}}}\right)\\
=p(s_{0})\prod_{t=0}^{T-1}\left(p(a_{t}\mid s_{t})\sum_{v_{t}}p(s_{t+1},v_{t}\mid s_{t},a_{t})\right)\\
=p(s_{0})\prod_{t=0}^{T-1}\left(p(a_{t}\mid s_{t})p(s_{t+1}\mid s_{t},a_{t})\right)\\
\end{array}
\] 

Thus, the only difference between a DEDP and an MDP is that a DEDP
introduces events to describe the state transition dynamics. Compared
with the aforementioned models introducing independence constraints
to make MDPs tractable, a DEDP does not make any independence assumptions.
It defines a simulation process that describes how components
interact and trigger an event, and how the aggregation of events
leads to system state changes. In this way, a DEDP captures the microscopic
dynamics in a macroscopic system, leading to more accurate dynamics
modeling.

\subsection{Duality Theorem on Value Function }

To solve a DEDP with a high-dimensional state and action space, we
derive the convex conjugate duality between the log expected future
reward function and the entropy function of a distribution over finite-length
DEDP trajectories, and the corresponding duality in the parameter
space between the log discounted trajectory-weighted reward and the
distribution of finite-length DEDP trajectories. As a result, we can
reduce the policy evaluation problem to a variational inference problem that
involves the entropy function and can be solved by various probabilistic
inference techniques. 

Specifically, in a complex system decision process\\ $\text{DEDP}\langle S,A,\mathcal{V},C,P,R,\gamma\rangle$,
let $T$ be a discrete time, $m$ the component index, $\xi_{T}$
the $\text{length-}T$ trajectory of a DEDP starting from initial
state $s_{0}$, and $V^{\pi}=E\left(\sum_{t=0}^{\infty}\gamma^{t}R_{t};\pi\right)$
the expected future reward (value function), where $\gamma\in[0,1)$
is a discount factor. Define $q(T,m,\xi_{T})$ as a proposal joint
probability distribution over finite $\text{length}-T$ DEDP trajectories,
$r(T,m,\xi_{T};\pi)=\gamma^{T}P(\xi_{T};\pi)R_{T}^{(m)}(s_{T}^{(m)})$
as the discounted trajectory-weighted reward component where $P(\xi_{T};\pi)$
is the probability distribution of a trajectory with policy $\pi$,
and $R_{T}^{(m)}(s_{T}^{(m)})$ as the reward evaluated at component
$m$ at time $T$ with state $s_{T}^{(m)}$. We thus have the following
duality theorem.

$\boldsymbol{Theorem\text{ }1.}$ \vspace{-4mm}

{\small{}
\[
\text{log}V^{\pi}(r)=\underset{q}{\text{sup}}\left(\sum_{T,m,\xi_{T}}q(T,m,\xi_{T})\text{log }r(T,m,\xi_{T};\pi)+\text{H}\left(q\right)\right)
\]
}{\small \par}

In the above, equality is satisfied when $\sum_{T}\sum_{m}\sum_{\xi_{T}}r(T,m,\xi_{T};\pi)<\infty$
and $q(T,m,\xi_{T})\propto\gamma^{T}P(\xi_{T})R_{T}^{(m)}(s_{T}^{(m)})$.
As such, $\text{log}V^{\pi}(r)$ is the convex conjugate of $\text{H}\left(q(T,m,\xi_{T})\right)$.
$\text{H}\left(q(T,m,\xi_{T})\right)$ is a convex function of $q(T,m,\xi_{T})$,
so by property of the convex conjugate, $\text{H}\left(q(T,m,\xi_{T})\right)$
is also a conjugate of $\text{log}V^{\pi}(r)$. The proof for this
theorem is shown in the Appendix.

Theorem 1 provides a tight lower bound of the log expected reward
and, more importantly, defines a variational problem in terms analogous
to well-known variational inference formulations in the graphic model
community \cite{wainwright2008graphical}, where a number of variational
inference methods can be introduced such as exact inference methods,
sampling-based approximate solutions, and variational inference. Theorem
1 extends the equivalence between optimal control and probability
inference in recent literatures to a general variational functional
problem. Specifically, it gets rid of the probability likelihood interpretation
of the value function in \cite{toussaint2010expectation,toussaint2006probabilistic},
and the prior assumption that value function is in the multiplication
form of local-scope value functions \cite{liu2012belief}.

\subsection{Expectation Propagation for Optimal Control}

Theorem 1 implies a generalized policy-iteration paradigm around the
duality form: solving the variational problem as policy evaluation
and optimizing the target over parameter $\theta$ with a known mixture
of finite-length trajectories as policy improvement. In the following,
we develop the policy evaluation and improvement algorithm with a
deterministic policy $a_{t}=\mu(s_{t};\theta)$, the derivation for
which is given in the Appendix. The stochastic policy case $\pi=p(a_{t}|s_{t};\theta)$
will lead to a similar result, which is not presented here due to the limit of space. 

In policy evaluation, the Markov property of a DEDP admits factorizations
$P(\xi_{T};\pi)=p(s_{0})\prod_{t=1}^{T}\left(p(v_{t}\mid s_{t};\theta)\delta_{s_{t+1}=s_{t}+\Delta_{v_{t}}}\right)$
and $q(T,m,\xi_{T})=q(T,m)q(s_{0})\prod_{t=1}^{T}q(s_{t-1,t},v_{t-1}|T)/\prod_{t=1}^{T-1}q(s_{t}|T)$,\\
where $q(T,m)=\gamma^{T}(1-\gamma)/M$ is the length prior distribution
to match the discount factor, and $q(s_{t}|T)$ and $q(s_{t-1,t},v_{t-1}|T)$
are locally consistent one-slice and two-slice marginals. To cope
with the exploding state space, we apply the Bethe entropy approximation.
Specifically, we relax the formidable searching state space of $s_{t}$
into an amenable space through the mean field approximation $q(s_{t}|T,m)=\prod_{\hat{m}=1}^{M}q(s_{t}^{(\hat{m})}|T,m)$,
where $q(s_{t}^{(\hat{m})}|T,m)$ is the one-slice marginal involving
only component $\hat{m}$. Applying the factorization and approximation,
we get the following Bethe entropy problem (let $\sum_{s_{t-1,t},v_{t-1}\setminus s_{t}^{(\hat{m})}}$
represent the summation over all value combinations of $s_{t-1},s_{t},v_{t-1}$
except a fixed $s_{t}^{(\hat{m})}$). \vspace{-4mm}

{\small{}
\begin{align}
 & \text{max over }q(s_{t}^{(\hat{m})}|T,m),q(s_{t-1,t},v_{t}|T,m)\text{ }\forall T,t\leq T,m\nonumber \\
 & \thinspace\sum_{T,m}\sum_{t=1}^{T-1}\sum_{\hat{m}}\sum_{s_{t}^{(\hat{m})}}\sum_{T,m}q(T,m,s_{t}^{(\hat{m})})\text{log}q(s_{t}^{(\hat{m})}|T,m)\nonumber \\
 & \begin{array}{l}
\thinspace-\sum_{T,m}\sum_{t=1}^{T-1}\sum_{s_{t-1,t},v_{t-1}}q(T,m,s_{t-1,t},v_{t-1})\text{log}\left(\frac{q(s_{t-1,t},v_{t-1}\mid T,m)}{p(s_{t},v_{t-1}\mid s_{t-1};\theta)}\right)\\
\thinspace\thinspace-\sum_{T,m,s_{T-1,T},v_{T-1}}q(T,m,s_{T-1,T},v_{T-1})\text{log}\left(\frac{q(s_{T-1,T},v_{T-1}\mid T,m)}{p(s_{T},v_{T-1}\mid s_{T-1};\theta)R_{T}^{(m)}}\right)
\end{array}\label{eq:problemafterapprox}\\
 & \text{subject to:}\begin{array}{ll}
\sum_{s_{t-1,t},v_{t-1}\setminus s_{t-1}^{(\hat{m})}}q(s_{t-1,t},v_{t-1}|T,m) & =q(s_{t-1}^{(\hat{m})}|T,m),\\
\sum_{s_{t-1,t},v_{t-1}\setminus s_{t}^{(\hat{m})}}q(s_{t-1,t},v_{t-1}|T,m) & =q(s_{t}^{(\hat{m})}|T,m)
\end{array}\nonumber 
\end{align}
}{\small \par}

We solve this with the method of Lagrange multipliers, which leads
to a forward-backward algorithm that updates the forward messages
$\alpha_{t|T,m}^{(\hat{m})}(s_{t}^{(\hat{m})})$ and backward messages
$\beta_{t|T,m}^{(\hat{m})}(s_{t}^{(\hat{m})})$\\ marginally according
to the average effects of all other components \textemdash{} that
is, a projected marginal kernel $p(s_{t}^{(\hat{m})},a_{t}|s_{t-1}^{(\hat{m})};\theta)$.
Therefore, the algorithm achieves linear complexity over the number
of components for each $T$ and $m$, and quadratic time complexity
over time horizon $H$ to compute all messages for all $T\le H$.
To further lower down the time complexity, we define ${\scriptstyle \text{\ensuremath{\beta_{t}^{(\hat{m})}}(\ensuremath{s_{t}^{(\hat{m})}})=\ensuremath{\underset{m}{\sum}\stackrel[T=t]{\infty}{\sum}}q(T,m)\ensuremath{\beta_{t|T,m}^{(\hat{m})}}(\ensuremath{s_{t}^{(\hat{m})}})}}$
by gathering together backward messages sharing $t$ and $\alpha_{t}^{(\hat{m})}(s_{t}^{(\hat{m})})=\alpha_{t|T,m}^{(\hat{m})}(s_{t}^{(\hat{m})})$
by noting that $\alpha_{t|T,m}^{(\hat{m})}(s_{t}^{(\hat{m})})$ doesn't
depend on $T,m$. This leads to the following forward-backward algorithm,
which is linear in time horizon: \vspace{-4mm}

\begin{align}
\alpha_{t}^{(\hat{m})}(s_{t}^{(\hat{m})}) & \propto\underset{s_{t-1}^{(\hat{m})},v_{t-1}}{\sum}\alpha_{t-1}^{(\hat{m})}(s_{t-1}^{(\hat{m})})\cdot p(s_{t}^{(\hat{m})},v_{t-1}|s_{t-1}^{(\hat{m})};\theta)\label{eq:forward-backward-alpha}\\
\beta_{t}^{(\hat{m})}(s_{t}^{(\hat{m})}) & =\underset{m}{\sum}q(t,m)\beta_{t|t,m}^{(\hat{m})}(s_{t}^{(\hat{m})})\nonumber \\
 & +\underset{s_{t+1}^{(\hat{m})},v_{t}}{\sum}p(s_{t+1}^{(\hat{m})},v_{t}|s_{t}^{(\hat{m})};\theta)\beta_{t+1}^{(\hat{m})}(s_{t+1}^{(\hat{m})})\label{eq:forward-backward-beta}
\end{align}

In policy improvement, we maximize the log expected future reward
function $L(\theta)=\sum_{T,m,\xi_{T}}q(T,m,\xi_{T};\theta^{\text{old}})\text{log}\left(\gamma^{T}P(\xi_{T};\theta)R_{T}^{(m)}\right)$
over parameter $\theta$ with $q(T,m,\xi_{T})$ inferred from $\theta^{\text{old}}$
via gradient ascent update $\left.\theta^{\text{new}}=\theta^{\text{old}}+\epsilon\cdot\frac{\partial L}{\partial\theta}\right|{}_{\theta^{\text{old}}}$,
or more aggressively by setting $\theta^{\text{new}}$ so that $\frac{\partial L}{\partial\theta}\left|_{\theta^{\text{new}}}=0\right.$.
This objective can be simplified by dropping irrelevant terms and
keeping only those involving policy: $L(\theta)=\underset{T,m,t,v_{t},s_{t}}{\sum}q(T,m,v_{t},s_{t};\theta^{\text{old}})\cdot\text{log}P(v_{t}|s_{t};\theta)+\text{const}$.
The gradient is obtained from chain rule and messages $\alpha_{t}(x_{t})$,
$\beta_{t}(x_{t})$ through dynamic programming: \vspace{-4mm}

\begin{align}
 & {\scriptstyle \frac{\partial L(\theta)}{\partial\theta}}={\scriptstyle \underset{t,s_{t}}{\sum}\frac{\prod_{\hat{m}}\alpha_{t}^{(\hat{m})}(s_{t}^{(\hat{m})},v_{t}=v)\beta_{t}^{(\hat{m})}(s_{t}^{(\hat{m})},v_{t}=v)}{c_{v}}\frac{\partial c_{v}}{\partial\theta}}\nonumber \\
 & {\scriptstyle -\underset{t,s_{t}}{\sum}\frac{\prod_{\hat{m}}\alpha_{t}^{(\hat{m})}(s_{t}^{(\hat{m})},v_{t}=\emptyset)\beta_{t}^{(\hat{m})}(s_{t}^{(\hat{m})},v_{t}=\emptyset)\cdot\prod_{m}g_{v}^{m}(s_{t}^{(m)})}{1-\sum\nolimits _{v=1}^{V}c_{v}\cdot\prod_{m}g_{v}^{m}(s_{t}^{(m)})}\frac{\partial c_{v}}{\partial\theta}}\label{eq:gradient-cv-1}
\end{align}

In summary, we give our optimal control algorithm of complex systems
as Algorithm 1. \vspace{-2mm}

\begin{algorithm}
\begin{flushleft}
\noindent\textbf{Input:} The DEDP tuple $\text{DEDP}\langle S,A,\mathcal{V},C,P,R,\gamma\rangle$, initial policy parameter $\theta$

\textbf{Output:} Optimal policy parameter $\theta$

\textbf{Procedure:} Iterate until convergence:

\begin{itemize}
\item policy evaluation
\end{itemize}
Iterate until convergence:

- Update forward messages according to Eq. (\ref{eq:forward-backward-alpha})

- Update backward messages according to Eq. (\ref{eq:forward-backward-beta})
\begin{itemize}
\item policy improvement
\end{itemize}
Update parameter $\theta$ according to the gradient in Eq. (\ref{eq:gradient-cv-1}).
\end{flushleft}
\caption{Optimal control of social systems\label{alg:Optimal-control-of}}
\end{algorithm}
\vspace{-4mm}

\subsection{Discussions}

In the above we developed a DEDP for modeling complex system decision
making problems, and derived a expectation propagation algorithm for
solving the DEDP. our algorithm is also applicable on a Markov decision
process with other simulation models. 
While we used a discounted expected total future reward in the previous derivation, our framework
is also applicable to other types of expected future reward, such
as a finite horizon future reward, where we use a different probability
distribution of time $q(T)=\frac{1}{T}$.

\section{Experiments}

In this experiment, we benchmark algorithm \ref{alg:Optimal-control-of}
against several decision-making algorithms for finding the optimal
policy in a complex system. 

\textbf{Overview:} The complex social system in this example is a transportation
system. The goal is to optimize the policy such that each vehicle
arriving at the correct facilities at correct time (being at work
during work hours and at home during rest hours) and spending minimum
time on roads. We formulate the transportation optimal control problem
as a discrete event decision process $\text{DEDP}\langle S,A,\mathcal{V},C,P,R,\gamma\rangle$.
The state variables $s_{t}=(s_{t}^{(1)},\cdots,s_{t}^{(M-1)},t)$
represent the populations at $(M-1)$ locations and the current time
$t$.  All events are of the form $p\cdot m_{1}{\scriptstyle \stackrel{c_{m_{1}m_{2}}}{\to}}p\cdot m_{2}$\textemdash an
individual $p$ moving from location $m_{1}$ to location $m_{2}$
with rate (probability per unit time) $c_{m_{1}m_{2}}$\textemdash decreasing
the population at $m_{1}$ by one and increasing the population at
$m_{2}$ by one. We also introduce auxiliary event $\emptyset$ that
doesn't change any system state, and set the rates of leaving facilities
and selecting alternative downstream links as action variables. We
implement the state transition $p(s_{t+1},v_{t}\mid s_{t},a_{t})$
following the fundamental diagram of traffic flow \cite{horni2016multi}
that simulate the movement of vehicles. The reward function $R(s_{t})=\sum_{m}\beta_{t,\text{perf}}^{(m)}s_{t}^{(m)}+\beta_{t,\text{trav}}^{(m)}s_{t}^{(m)}$
emulates the Charypa-Nagel scoring function in transportation research
\cite{horni2016multi} to reward performing the correct activities
at facilities and penalize traveling on roads , where $\beta_{t,\text{trav}}^{(m)}$
and $\beta_{t,\text{perf}}^{(m)}$ are the score coefficients. We
implement the deterministic policy as a function of states through
a neural network $a_{t}=\mu(s_{t})=\mathcal{NN}(s_{t};\theta)$ and
apply Algorithm 1 to find the optimal policy parameter $\theta$. 

\textbf{Benchmark Model Description:}  The transition kernel based
on an MDP in this general form $p(s_{t+1}\mid s_{t},a_{t})$ in a complex
system is too complicated to be modeled exactly with an analytical
form, due to the high-dimensional state-action space, and the complex
system dynamics. As such, we benchmark with the following algorithms:
(1) Analytical approaches based on an MDP that uses Taylor expansion
to approximate the intractable transition dynamics with differential
equations, which leads to a guided policy search (GPS) algorithm \cite{montgomery2016guided}
with a known dynamics that uses iterative linear quadratic regulator
(iLQR) for trajectory optimization and supervised learning for training
a global policy $\mu(s_{t})$, implemented as a five-layer neural
network. Other aforementioned approximations \cite{peyrard2006mean,sabbadin2012framework,cheng2013variational}
are not applicable because in their settings each component takes
an action, resulting in local policies for each component. Whereas
in our problem the action is taken by the system as a whole, and therefore
no local policies. (2)  Simulation approaches that reproduce the
dynamic flow through sampling the state action trajectories from the
current policy and the system transition dynamics,  which leads to
a policy gradient (PG) algorithm, the policy of which
is implemented as a four-layer neural network; and (3) an actor-critic
(AC) algorithm \cite{lillicrap2015continuous} that implements the
policy $\mu(s_{t})$ as an actor network with four layers and the
state-action value function $Q(s_{t},a_{t})$ as a critic network
with five layers. 

\begin{figure*}[t]
\hspace{-.5in}
\begin{minipage}[t]{0.48\columnwidth}%
\begin{figure}[H]
\begin{centering}
\includegraphics[width=0.28\paperwidth,height=0.14\paperheight]{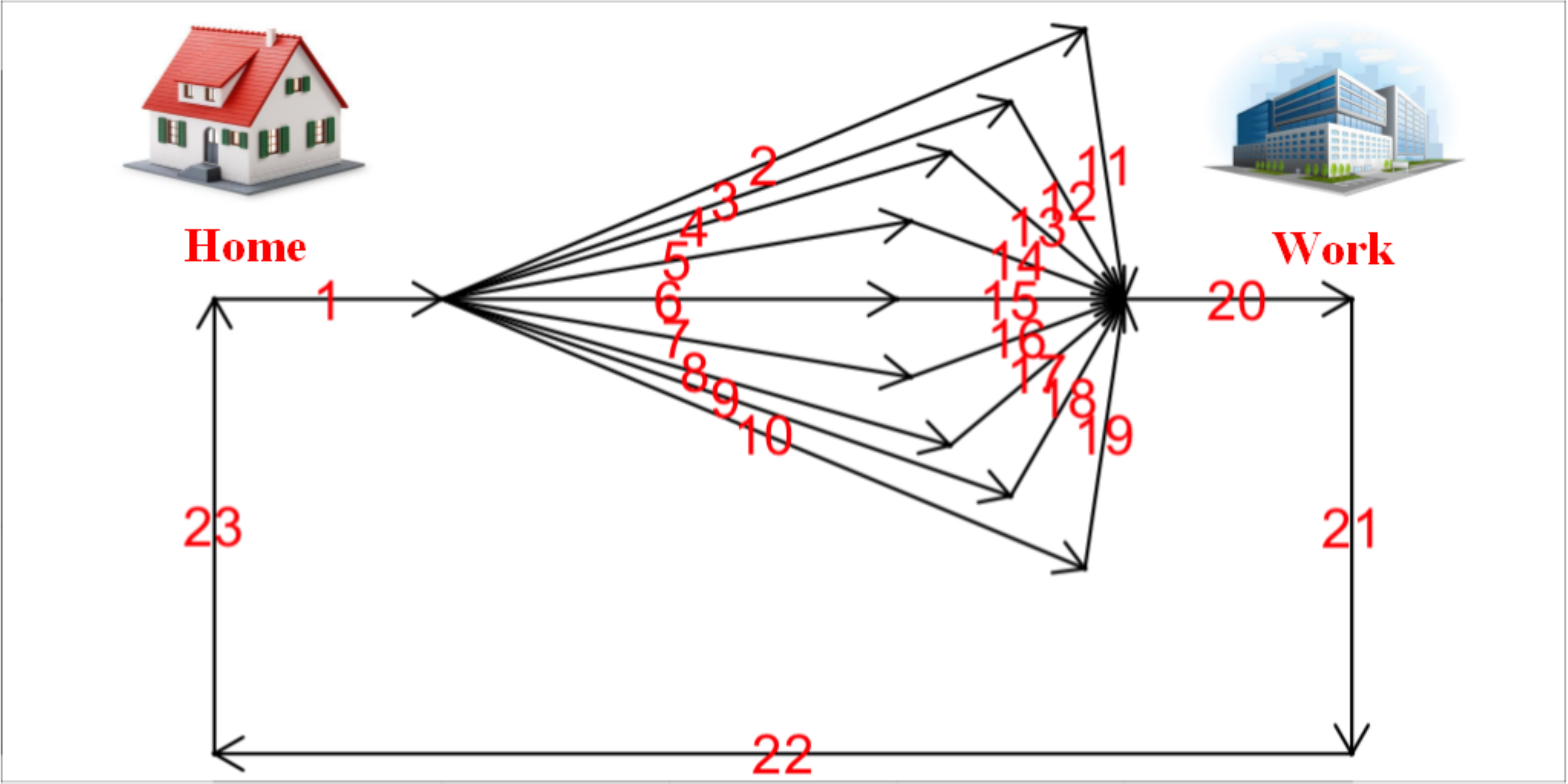}
\par\end{centering}
\caption{SynthTown road network\label{fig:SynthTownMap}}
\vspace{-4mm}
\end{figure}
\end{minipage}\qquad{}\qquad{}\qquad{}\quad{}%
\begin{minipage}[t]{0.48\columnwidth}%
\begin{figure}[H]
\begin{centering}
\includegraphics[width=0.25\paperwidth,height=0.15\paperheight]{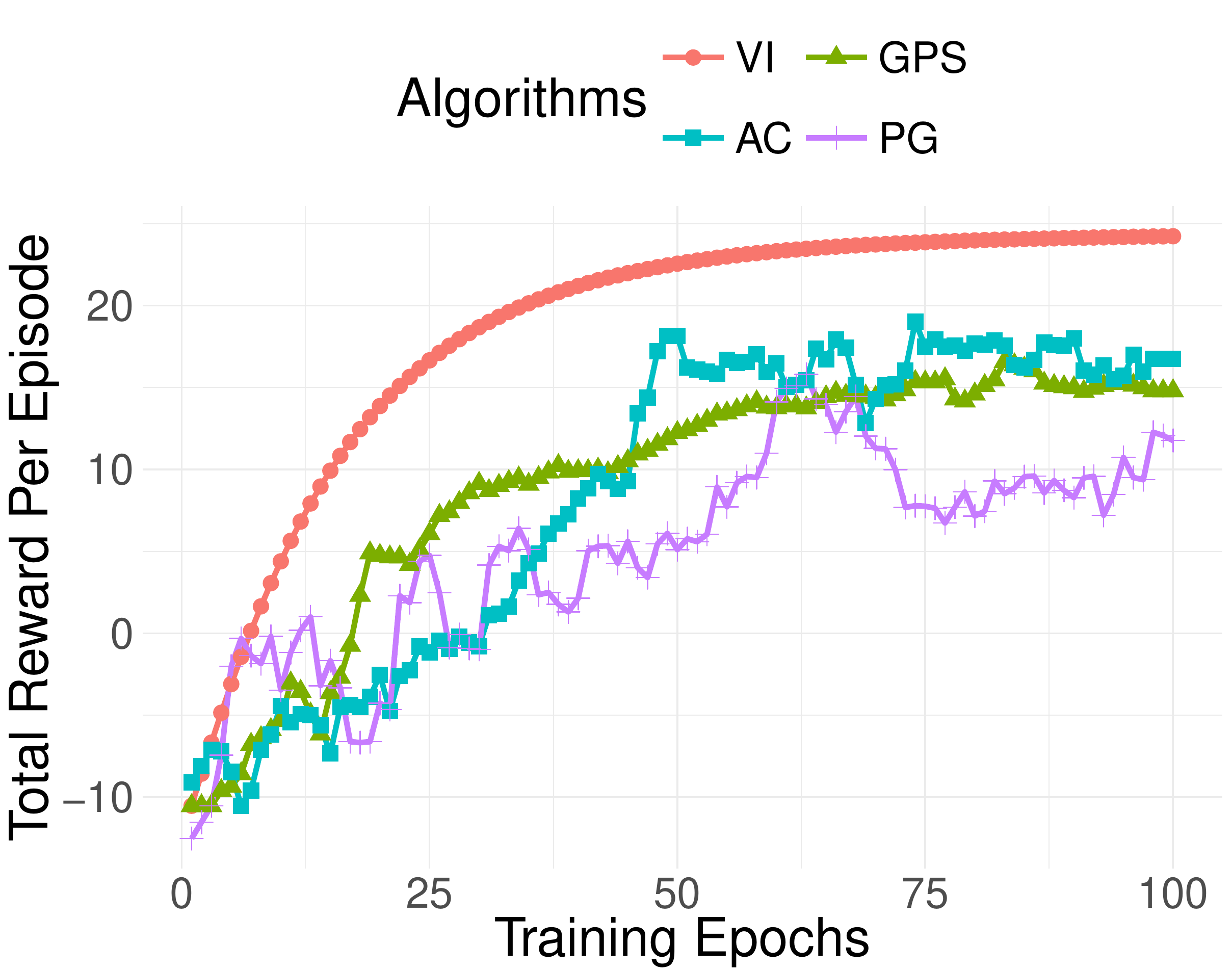}
\par\end{centering}

\caption{Training process on SynthTown\label{fig:SynthTown_value_epochs}}
\vspace{-4mm}
\end{figure}
\end{minipage}\qquad{}\qquad{}\quad{}%
\noindent\begin{minipage}[t]{0.51\columnwidth}%
\begin{figure}[H]
\begin{centering}
\includegraphics[clip,width=0.25\paperwidth,height=0.15\paperheight]{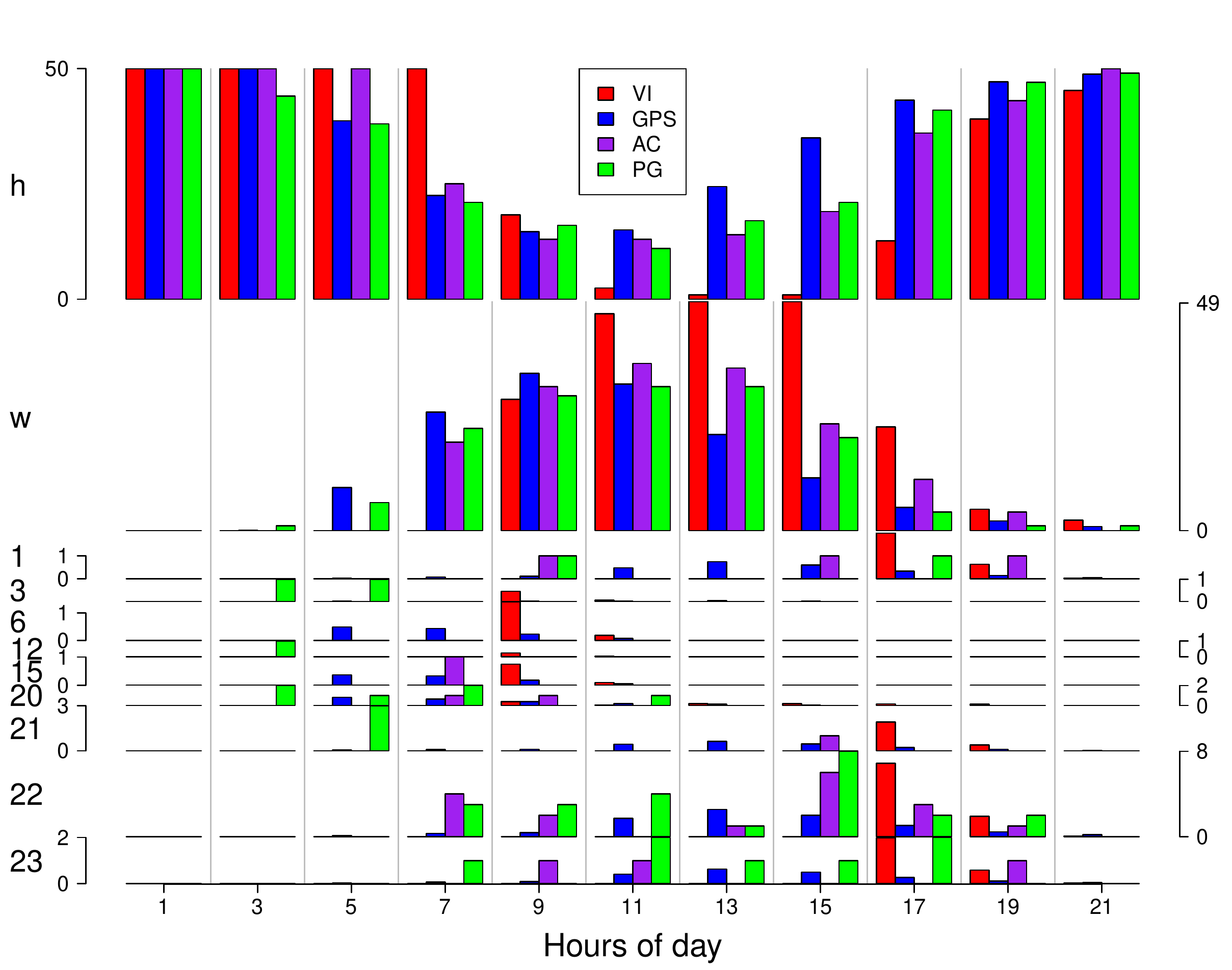}
\par\end{centering}

\caption{Average number of vehicles with trained policies\label{fig:SynthTown_xt_t}}
\vspace{-4mm}
\end{figure}
\end{minipage}

\end{figure*}

\textbf{Performance and Efficiency:} We benchmark the algorithms
using the SynthTown scenario (Fig. \ref{fig:SynthTownMap}), which
has one home facility, one work facility, 23 road links, and $50$
individuals going to work facility in the morning (9 am) and home
facility in the evening (5 pm).  A training episode is one day. This
scenario is small enough for studying the details of different algorithms.

In Figure \ref{fig:SynthTown_value_epochs}, the value-epoch curve
of VI (our algorithm) dominates those of the other algorithms almost
everywhere. Table \ref{tab:StatisticsTable} indicates that VI requires
the fewest training epochs to converge to the highest total rewards
per episode. Figure \ref{fig:SynthTown_xt_t} presents the average
vehicle distribution of ten runs at different locations (Home (h),
Work (w), and roads 1-23) using the learned policy with each algorithm. This figure implies that the learned policy of VI leads to the largest
amount of vehicles being at work during work hours (9 am to 5 pm),
and least amount of time the vehicles spending on roads.

In the SynthTown scenario, high rewards require the individuals to perform
the correct activities (home, work, and so on) at the right time,
and to spend less time on roads. VI achieves the best performance
when evaluating a policy by considering the whole state-action space
with VI approximation \textemdash{} the evolution of each state variables
in the mean field of the other state variables. Modeling the complex
system transition dynamics based on an MDP analytically with Taylor
approximation will introduce modeling errors (GPS). Value estimation
from Monte Carlo integration in high-dimensional state space has high
variance (PG). A small perturbation of policy will result in significant
change to the immediate reward and value in later steps, which makes
it difficult to estimate the correct value from sampled trajectories,
and as a result difficult to compute the correct value gradient (AC).

\begin{table}[H]
\begin{centering}
{\small{}}%
\begin{tabular}{|c||c|c|c|c|}
\hline 
{\small{}Dataset} & \multicolumn{2}{c|}{{\small{}SynthTown}} & \multicolumn{2}{c|}{{\small{}Berlin}}\tabularnewline
\hline 
{\small{}Metrics} & \textbf{TRPE} & \textbf{EC} & \textbf{TRPE} & \textbf{EC}\tabularnewline
\hline 
\hline 
VI & 24.21 & 75 & 11.52 & 200\tabularnewline
\hline 
GPS & 14.79 & 100 & -10.33  & -\tabularnewline
\hline 
AC & 16.76 & 100 & -9.69  & -\tabularnewline
\hline 
PG & {\small{}11.77} & 150 & -15.52 & -\tabularnewline
\hline 
\end{tabular}
\par\end{centering}{\small \par}
\caption{Comparing algorithms in total reward per episode (TRPE) and epochs
to converge (EC) \label{tab:StatisticsTable}}
\end{table}
\vspace{-4mm}

We also benchmark the performance of all algorithms using the Berlin
scenario, which consists of a network of 1,530 locations and the trips
of $9,178$ synthesized individuals \cite{horni2016multi}. Table
\ref{tab:StatisticsTable} shows that VI outperformed in total rewards
per episode, while the other algorithms did not even converge in a
reasonable number of epochs. 

In summary, VI outperformed guided policy search, policy gradient,
and actor-critic algorithms in all scenarios. These benchmarking algorithms
provide comparable results in a small dataset such as the SynthTown
scenario, but became difficult to train when applied to a larger dataset
such as Berlin. \vspace{4mm}

\section{Related Works}
A number of prior works have explored the connection between decision
making and probabilistic inference. The earliest such research is
the Kalman duality proposed by Rudolf Kalman \cite{kalman1960new}.
Subsequent works have expanded the duality in stochastic optimal control,
MDP planning, and reinforcement learning. Todorov and Kappen expanded
the connection between control and inference by restricting the immediate
cost and identifying the stochastic optimal control problems as linearly-solvable
MDPs \cite{todorov2007linearly} or KL control problems \cite{kappen2012optimal}.
However, the passive dynamics in a linearly-solvable MDPs and the
uncontrolled dynamics in a KL control problem become intractable 
 in a social system scenario due to the complicated system dynamics.
Toussaint, Hoffman, David, and colleagues broadened the connection
by framing planning in an MDP as inference in a mixture of PGMs \cite{toussaint2006probabilistic}.
The exact \cite{furmston2012efficient} and variational inference
\cite{furmston2010variational} approaches in their framework encounter
the transition dynamics intractability issue in a social network setting
due to the complicated system dynamics. Their sampling-based alternatives
\cite{hoffman2009new} experience the high-variance problem in a social
network due to the exploding state space. Levine, Ziebart, and colleagues
widened the connection by establishing the equivalence between maximum
entropy reinforcement learning, where the standard reward objective
is augmented with an entropy term \cite{ziebart2010modeling}, and
probabilistic inference in a PGM \cite{ziebart2010modeling,levine2018reinforcement}.
They optimize a different objective function compared with our method.
Our approach is an extension of Toussaint's formulation \cite{toussaint2006probabilistic},
but differs in that we establish the duality with a DEDP, and provide
new insights into the duality from the perspective of convex conjugate
duality. Compared with the existing exact inference \cite{furmston2012efficient}
and approximate inference solutions \cite{hoffman2009new,furmston2010variational},
our algorithm is more efficient with gathering the messages, and more
scalable with the use of Bethe entropy approximation. 

Other formulations of the decision-making problems which are similar
to our framework are the Option network \cite{sutton1999between} and the multi-agent Markov decision process (MMDP)
\cite{sigaud2013markov}.
An option network extends a traditional Markov decision process with options --- closed loop policies for taking actions over a period of time, with the goal of providing a temporal abstraction of the representation. In our framework, we introduce an auxiliary variable $v$ with the goal of providing a more accurate and efficient modeling of the system dynamics. 
An MMDP represents sequential decision-making
problems in cooperative multi-agent settings. There are two fundamental
differences between our framework and MMDP. First, unlike MMDP where
each agent takes an action at each time step, in a DEDP the actions
are taken by the system and the dimensionality of states not necessarily
equals to the dimensionality of actions. Second, unlike MMDP
where each agent has its own information, a DEDP models the decision
making of a large system where only one controller has all the information
and makes decisions for the entire system.

\section{Conclusions}

In this paper, we have developed DEDP for modeling complex system decision-making
processes. We reduce the optimal control of DEDP to variational inference
and parameter learning around a variational duality theorem. Our framework
is capable of optimally controlling real-world complex systems, as
demonstrated by experiments in a transportation scenario.



\section{Appendix}

\subsection{Derivation of Theorem 1}

\[
\begin{array}{ll}
{\scriptstyle \ensuremath{\text{log}V^{\pi}}} & {\scriptstyle \ensuremath{=\text{log}\left(\sum_{T}\sum_{m}\sum_{\xi_{T}}q(T,m,\xi_{T})\frac{r(T,m,\xi_{T};\pi)}{q(T,m,\xi_{T})}\right)}}\\
 & {\scriptstyle \ensuremath{\geq\sum_{T}\sum_{m}\sum_{\xi_{T}}q(T,m,\xi_{T})\text{log}\frac{r(T,m,\xi_{T};\pi)}{q(T,m,\xi_{T})}}}\\
 & {\scriptstyle \ensuremath{=\sum_{T}\sum_{m}\sum_{\xi_{T}}q(T,m,\xi_{T})\text{log }r(T,m,\xi_{T};\pi)+\text{H}\left(q(T,m,\xi_{T})\right)}}
\end{array}
\]

\subsection{Derivation of Eq. (\ref{eq:problemafterapprox})}

Rearranging the terms and applying the approximations described in
the main text, the target becomes the following:\vspace{-5mm}

\[
\begin{array}{ll}
 & {\scriptstyle \ensuremath{\sum_{T}\sum_{m}\sum_{\xi_{T}}q(T,m,\xi_{T})\text{log}\left(\gamma^{T}P(\xi_{T};\theta)R_{T}^{(m)}\right)+\text{H}\left(q(T,m,\xi_{T})\right)}}\\
 & {\scriptstyle \ensuremath{=\sum_{T}\sum_{m}\sum_{\xi_{T}}q(T,m,\xi_{T})\text{log}\left(\gamma^{T}\stackrel[t=1]{T-1}{\prod}p(s_{t},v_{t-1}\mid s_{t-1};\theta)\cdot p(s_{T},v_{T-1}\mid s_{T-1};\theta)R_{T}^{(m)}\right)}}\\
 & {\scriptstyle \ensuremath{-\sum_{T}\sum_{m}\sum_{\xi_{T}}q(T,m,\xi_{T})\text{log}\left(q(T,m)\frac{\stackrel[t=1]{T}{\prod}q(s_{t-1,t},v_{t-1}\mid T,m)}{\stackrel[t=1]{T-1}{\prod}q(s_{t}\mid T,m)}\right)}}\\
 & {\scriptstyle {\scriptscriptstyle =\sum_{T}\sum_{m}(q(T,m)\text{log}\left(\frac{\gamma^{T}}{q(T,m)}\right)-\stackrel[t=1]{T-1}{\sum}\sum_{s_{t-1,t},v_{t-1}}q(T,m,s_{t-1,t})\text{log}\left(\frac{q(s_{t-1,t},v_{t-1}\mid T,m)}{p(s_{t},v_{t-1}\mid s_{t-1};\theta)}\right)}}\\
 & {\scriptstyle \ensuremath{\text{ }-\sum_{t}\sum_{m}\sum_{s_{t-1,t},v_{t-1}}q(T,m,s_{t-1,t})\text{log}\left(\frac{q(s_{t-1,t},v_{t-1}\mid T,m)}{p(s_{t},v_{t-1}\mid s_{t-1};\theta)R_{t}^{(m)}}\right)}}\\
 & {\scriptstyle \ensuremath{+\sum_{T,m}\stackrel[t=1]{T-1}{\sum}\sum_{\hat{m}}\sum_{s_{t}^{(\hat{m})}}q(T,m,s_{t}^{(\hat{m})})\text{log}q(s_{t}^{(\hat{m})}|T,m)}}
\end{array}
\]

\subsection{Derivation to Solve Eq. (\ref{eq:problemafterapprox})}

We solve this maximization problem with the method of Lagrange multipliers.
\vspace{-5mm}

\[
\begin{array}{ll}
{\scriptstyle \ensuremath{L}} & {\scriptstyle \ensuremath{=\sum_{T,m}q(T,m)\text{log}\left(\frac{\gamma^{T}}{q(T,m)}\right)}}\\
 & {\scriptstyle \ensuremath{-\sum_{T,m}\stackrel[t=1]{T-1}{\sum}\sum_{s_{t-1,t},v_{t-1}}q(T,m,s_{t-1,t},v_{t-1})\text{log}\left(\frac{q(s_{t-1,t},v_{t-1}\mid T,m)}{p(s_{t},v_{t-1}\mid s_{t-1};\theta)}\right)}}\\
 & {\scriptstyle \ensuremath{-\sum_{T,m}\sum_{s_{T-1,T}}q(T,m,s_{T-1,T},v_{T-1})\text{log}\left(\frac{q(s_{T-1,T},v_{T-1}\mid T,m)}{p(s_{T},v_{T-1}\mid s_{T-1};\theta)R_{T}^{(m)}}\right)}}\\
 & {\scriptstyle \ensuremath{+\sum_{T,m}\stackrel[t=1]{T-1}{\sum}\sum_{\hat{m}}\sum_{s_{t}^{(\hat{m})}}q(T,m,s_{t}^{(\hat{m})})\text{log}q(s_{t}^{(\hat{m})}|T,m)}}\\
 & \hspace{-.2in}{\scriptstyle \ensuremath{+\sum_{t,\hat{m},s_{t-1}^{(\hat{m})},T,m}\alpha_{t-1,s_{t-1}^{(\hat{m})},T,m}^{\hat{m}}\left(\sum_{s_{t-1,t},v_{t-1}\setminus s_{t-1}^{(\hat{m})}}q(s_{t-1,t},v_{t-1}|T,m)-q(s_{t-1}^{(\hat{m})}|T,m)\right)}}\text{ }\\
 & \hspace{-.2in}{\scriptstyle \ensuremath{\text{ }+\sum_{t,\hat{m},s_{t}^{(\hat{m})},T,m}\beta_{t,s_{t}^{(\hat{m})},T,m}^{\hat{m}}\left(\sum_{s_{t-1,t},v_{t-1}\setminus s_{t}^{(\hat{m})}}q(s_{t-1,t},v_{t-1}|T,m)-q(s_{t}^{(\hat{m})}|T,m)\right)}}
\end{array}
\]

Taking derivative with respect to ${\scriptstyle q(s_{t-1,t},a_{t}\mid T,m)}$
and ${\scriptstyle q(s_{t}^{(\hat{m})}\mid T,m)}$, and setting it
to zero, we then get\vspace{-5mm}

\[
\begin{array}{ll}
{\scriptstyle \ensuremath{\text{for }t=1,...T-1}}\\
{\scriptstyle \ensuremath{q(s_{t-1,t},v_{t-1}|T,m)}} & {\scriptstyle \propto\ensuremath{\text{exp}\left(\frac{\sum_{\hat{m}}\alpha_{t-1,s_{t-1}^{(\hat{m})},T,m}^{(\hat{m})}}{q(T,m)}\right)\cdot p(s_{t},v_{t-1}\mid s_{t-1};\theta)\text{exp}\left(\frac{\sum_{\hat{m}}\beta_{t,s_{t}^{(\hat{m})},T,m}^{(\hat{m})}}{q(T,m)}\right)}}\\
{\scriptstyle \ensuremath{\text{ }\text{for }t=T}}\\
{\scriptstyle \ensuremath{q(s_{t-1,t},v_{t-1}|T,m)}} & {\scriptstyle \propto\ensuremath{\text{exp}\left(\frac{\sum_{\hat{m}}\alpha_{t-1,s_{t-1}^{(\hat{m})},T,m}^{(\hat{m})}}{q(T,m)}\right)\cdot p(s_{t},v_{t-1}\mid s_{t-1};\theta)R_{T}^{(m)}\text{exp}\left(\frac{\sum_{\hat{m}}\beta_{t,s_{t}^{(\hat{m})},T,m}^{(\hat{m})}}{q(T,m)}\right)}}\\
{\scriptstyle q(s_{t}^{(\hat{m})}\mid T,m)} & {\scriptstyle \ensuremath{=\frac{1}{Z_{T}^{(m)}}\text{exp}\left(\frac{\alpha_{t,s_{t}^{(\hat{m})},T,m}^{(\hat{m})}+\beta_{t,s_{t}^{(\hat{m})},T,m}^{(\hat{m})}}{q(T,m)}\right)}}\\
\\
\end{array}
\]

Marginalizing over ${\scriptstyle \ensuremath{q(s_{t-1,t}\mid T,m)}}$,
we have 

\[ \hspace{-.35in}
\begin{array}{ll}
 & {\scriptstyle \ensuremath{q(s_{t-1,t}^{(\hat{m})},v_{t-1}\mid T,m)}}\\
 & {\scriptstyle \ensuremath{=\sum_{s_{t-1,t}^{(\hat{m}')};\hat{m}'\neq\hat{m}}\frac{1}{Z_{t}}\text{exp}\left(\frac{\sum_{\hat{m}}\alpha_{t-1,s_{t-1}^{(\hat{m})},T,m}^{(\hat{m})}}{q(T,m)}\right)\cdot p(s_{t},v_{t-1}\mid s_{t-1};\theta)\text{exp}\left(\frac{\sum_{\hat{m}}\beta_{t,s_{t}^{(\hat{m})},T,m}^{(\hat{m})}}{q(T,m)}\right)}}\\
 & {\scriptstyle \ensuremath{:=\frac{1}{Z_{t}}\text{exp}\left(\frac{\alpha_{t-1,s_{t}^{(\hat{m})},T,m}^{(\hat{m})}+\beta_{t,s_{t}^{(\hat{m})},T,m}^{(\hat{m})}}{q(T,m)}\right)\cdot p(s_{t}^{(\hat{m})},v_{t-1}\mid s_{t-1}^{(\hat{m})};\theta)}}
\end{array}
\]

\hspace{-.35in}We denote ${\scriptstyle \ensuremath{\text{exp}\left(\frac{\alpha_{t,s_{t}^{(\hat{m})},T,m}^{(\hat{m})}}{q(T,m)}\right)}}$
as ${\scriptstyle \ensuremath{\alpha_{t|T,m}^{(\hat{m})}\left(s_{t}^{(\hat{m})}\right)}}$,
${\scriptstyle \ensuremath{\text{exp}\left(\frac{\beta_{t,s_{t}^{(\hat{m})},T,m}^{(\hat{m})}}{q(T,m)}\right)}}$
as ${\scriptstyle \ensuremath{\beta_{t|T,m}^{(\hat{m})}\left(s_{t}^{(\hat{m})}\right)}}$.

We can compute $\alpha,\,\beta$ through a forward-backward iterative
approach: 

\[\hspace{-3mm}
\begin{array}{ll}
{\scriptstyle \ensuremath{\text{forward:}}} & {\scriptstyle \ensuremath{q(s_{t}^{(\hat{m})}\mid T,m)=\sum_{s_{t-1}^{(\hat{m})},v_{t-1}}q(s_{t-1,t}^{(\hat{m})},v_{t-1}\mid T,m)}}\\
{\scriptstyle \ensuremath{\Rightarrow}} & \hspace{-.3in}{\scriptstyle \ensuremath{\alpha_{t|T,m}^{(\hat{m})}(s_{t}^{(\hat{m})})=\frac{Z_{t}^{(\hat{m})}}{Z_{t}}\sum_{s_{t-1}^{(\hat{m})},v_{t-1}}\alpha_{t-1|T,m}^{(\hat{m})}(s_{t-1}^{(\hat{m})})\cdot p(s_{t}^{(\hat{m})},v_{t-1}\mid s_{t-1}^{(\hat{m})};\theta)}}\\
{\scriptstyle \ensuremath{\text{backward:}}} & {\scriptstyle \ensuremath{q(s_{t-1}^{(\hat{m})}\mid T,m)=\sum_{s_{t}^{(\hat{m})},v_{t-1}}q(s_{t-1,t}^{(\hat{m})},v_{t-1}\mid T,m)}}\\
{\scriptstyle \ensuremath{\Rightarrow}} & \hspace{-.3in}{\scriptstyle \ensuremath{\beta_{t-1|T,m}^{(\hat{m})}(s_{t-1}^{(\hat{m})})=\frac{Z_{t}^{(\hat{m})}}{Z_{t}}\sum_{s_{t}^{(\hat{m})},v_{t-1}}p(s_{t}^{(\hat{m})},v_{t-1}\mid s_{t-1}^{(\hat{m})};\theta)\cdot\beta_{t|T,m}^{(\hat{m})}(s_{t}^{(\hat{m})})}}\\
 & {\scriptstyle \ensuremath{\text{for }t=T,\hat{m}=m}}\\
 & \hspace{-.3in}{\scriptstyle \ensuremath{\beta_{t-1|T,m}^{(\hat{m})}(s_{t-1}^{(\hat{m})})=\frac{Z_{t}^{(\hat{m})}}{Z_{t}}\sum_{s_{t}^{(\hat{m})},v_{t-1}}p(s_{t}^{(\hat{m})},v_{t-1}\mid s_{t-1}^{(\hat{m})};\theta)\cdot\beta_{t|T,m}^{(\hat{m})}(s_{t}^{(\hat{m})})R_{T}^{(m)}}}
\end{array}
\]

\subsection{Derivation of Eq. (\ref{eq:forward-backward-beta})}

 We can further simplify our algorithm by gathering the messages.
For notational simplicity, we can absorb the reward ${\scriptstyle \ensuremath{R_{T}^{(m)}(s_{T}^{(m)})}}$
into the ${\scriptstyle \ensuremath{\beta_{T|T,m}^{(m)}(s_{T}^{(m)})}}$
term, so that ${\scriptstyle \ensuremath{\beta_{T|T,m}^{(m)}(s_{T}^{(m)})=R_{T}^{(m)}(s_{T}^{(m)})}}$
and ${\scriptstyle \ensuremath{\beta_{T|T,m}^{(\hat{m})}(s_{T}^{(\hat{m})})=1}}$
for ${\scriptstyle \ensuremath{\hat{m}\neq m}}$. We then define ${\scriptstyle \ensuremath{\beta_{t}^{(\hat{m})}(s_{t}^{(\hat{m})})=\underset{m}{\sum}\stackrel[T=t]{\infty}{\sum}q(T,m)\beta_{t|T,m}^{(\hat{m})}(s_{t}^{(\hat{m})})}}$
and have \vspace{-5mm}

{\small{}
\[
\begin{array}{ll}
 & {\scriptstyle \ensuremath{\beta_{t}^{(\hat{m})}(s_{t}^{(\hat{m})})}}\\
 & {\scriptstyle \ensuremath{=\underset{m}{\sum}\stackrel[T=t]{\infty}{\sum}q(T,m)\beta_{t|T,m}^{(\hat{m})}(s_{t}^{(\hat{m})})}}\\
 & {\scriptstyle \ensuremath{=\underset{m}{\sum}q(t,m)\beta_{t|t,m}^{(\hat{m})}(s_{t}^{(\hat{m})})+\sum_{s_{t+1}^{(\hat{m})},v_{t}}p(s_{t+1}^{(\hat{m})},v_{t}\mid s_{t}^{(\hat{m})};\theta)\beta_{t+1}^{(\hat{m})}(s_{t+1}^{(\hat{m})})}}
\end{array}
\]
}{\small \par}

Observing that ${\scriptstyle \ensuremath{\stackrel[T=1]{\infty}{\sum}\stackrel[t=1]{T}{\sum}\iff\stackrel[t=1]{\infty}{\sum}\stackrel[T=t]{\infty}{\sum}}}$,
this summation form of messages will be useful in the parameter-learning
phase. 


\subsection{Derivation of Eq. (\ref{eq:gradient-cv-1})}

\begin{align*}
 & {\scriptstyle \ensuremath{\frac{\partial L}{\partial\theta}=\underset{T,m,t}{\sum}\underset{s_{t}}{\sum}\frac{q(T,m,v_{t}=v,s_{t})}{c_{v}}\frac{\partial c_{v}}{\partial\theta}}}\\
 & {\scriptstyle \ensuremath{\thinspace\thinspace-\underset{T,m,t}{\sum}\underset{s_{t}}{\sum}\frac{q(T,m,v_{t}=\emptyset,s_{t})\cdot g_{v}(s_{t})}{1-\sum\nolimits _{v=1}^{V}c_{v}g_{v}(s_{t})}\frac{\partial c_{v}}{\partial\theta}}}\\
 & \begin{array}{l}
{\scriptstyle \ensuremath{=\underset{t,s_{t}}{\sum}\frac{\prod_{\hat{m}}\alpha_{t}^{(\hat{m})}(s_{t}^{(\hat{m})},v_{t}=v)\beta_{t}^{(\hat{m})}(s_{t}^{(\hat{m})},v_{t}=v)}{c_{v}}\frac{\partial c_{v}}{\partial\theta}}}\\
{\scriptstyle \ensuremath{-\underset{t,s_{t}}{\sum}\frac{\prod_{\hat{m}}\alpha_{t}^{(\hat{m})}(s_{t}^{(\hat{m})},v_{t}=\emptyset)\beta_{t}^{(\hat{m})}(s_{t}^{(\hat{m})},v_{t}=\emptyset)\cdot\prod_{m}g_{v}^{m}(s_{t}^{(m)})}{1-\sum\nolimits _{v=1}^{V}c_{v}\cdot\prod_{m}g_{v}^{m}(s_{t}^{(m)})}\frac{\partial c_{v}}{\partial\theta}}}
\end{array}
\end{align*}



\newpage

\bibliographystyle{ACM-Reference-Format}  
\balance
\bibliography{reference}  

\end{document}